\documentclass[aps, prb, superscriptaddress, citeautoscript, showpacs, twocolumn]{revtex4-1}
\usepackage{float}
\usepackage[pdftex]{graphicx}
\usepackage{epstopdf}
\usepackage{amsmath}
\usepackage{subfigure}
\usepackage{fixltx2e}
\usepackage{dcolumn}
\begin{document}

\title{Evolution of structure and magnetism across the metal-insulator transition \protect\\ in the pyrochlore iridate $($Nd$_{1-x}$Ca$_x)_2$Ir$_2$O$_7$}

\author{Zach Porter}
\author{Eli Zoghlin}
\affiliation{Materials Department, University of California, Santa Barbara, California 93106, USA}

\author{Samuel Britner}
\author{Samra Husremovic}
\affiliation{Department of Chemistry and Biochemistry, Bates College, Lewiston, Maine 04240, USA}

\author{\\Jacob P. C. Ruff}
\affiliation{CHESS, Cornell University, Ithaca, New York 14853, USA}

\author{Yongseong Choi}
\author{Daniel Haskel}
\affiliation{Advanced Photon Source, Argonne National Laboratory, Argonne, Illinois 60439, USA}

\author{Geneva Laurita}
\affiliation{Department of Chemistry and Biochemistry, Bates College, Lewiston, Maine 04240, USA}

\author{Stephen D. Wilson}
\affiliation{Materials Department, University of California, Santa Barbara, California 93106, USA}
\email[email: ]{stephendwilson@ucsb.edu}

\date{\today}

\begin{abstract}
We report on the evolution of the thermal metal-insulator transition in polycrystalline samples of Nd$_2$Ir$_2$O$_7$ upon hole-doping via substitution of Ca$^{2+}$ for Nd$^{3+}$. Ca substitution mediates a filling-controlled Mott-like transition with minimal resolvable structural changes and without altering site symmetry. Local structure confirms that Ca substitution does not result in local chemical phase separation, and absorption spectroscopy establishes that Ir cations maintain a spin-orbit entangled electronic configuration. The metal-insulator transition coincides with antiferromagnetic ordering on the Ir sublattice for all measured samples, and both decrease in onset temperature with Ca content. Weak low-temperature upturns in susceptibility and resistivity for samples with high Ca content suggest that Nd sublattice antiferromagnetism continues to couple to carriers in the metallic regime.
\end{abstract}

\maketitle
\section{Introduction}

The pyrochlore structure A$_2$B$_2$O$_7$ is comprised of interpenetrating A and B sublattices of corner-sharing tetrahedra. This structure hosts a variety of electronic and magnetic phases owing in part to geometric frustration in the presence of antiferromagnetic exchange interactions \cite{Gardner2010} and the diversity of cation species which can be accommodated within this structural framework. In particular, setting B to Ir$^{4+}$ and A to a trivalent lanthanide ion (or Y$^{3+}$) results in a material with small energy gaps between quadratic bands. Upon varying the lanthanide site, the  A$_2$Ir$_2$O$_7$ series exhibits a metal-insulator transition (MIT) \cite{Matsuhira2007} where the gap monotonically decreases with increasing ionic radius \cite{Matsuhira2011} until a metallic state is reached between A=Nd and A=Pr lanthanide ions \cite{Ueda2015}. Interestingly, the transition from a metal into an insulator for the pyrochlore iridates coincides with the formation of all-in-all-out (AIAO) antiferromagnetic order of the $J$\textsubscript{eff}=1/2 moments on the Ir magnetic sublattice, while magnetic A site lanthanide cations establish the same order at lower temperatures. 

Theoretical studies predict that the suppression of both the insulating and magnetic phases results in a novel variety of antiferromagnetic quantum critical points (QCPs) and the formation of topologically nontrivial electronic states \cite{Wan2011,Savary2012,Witczak2012,Savary2014,Kimchi2014}.
Neighboring the quantum critical regime, the Weyl semimetal phase has been directly observed \cite{Ueda2018}, as have unconventional electronic properties nearby \cite{Sushkov2015,Fujita2015,Ueda2016}. However, fully exploring the experimental manifestation of this QCP has been challenging for the A$_2$Ir$_2$O$_7$ materials system, primarily because of ambiguities regarding the mechanism of the MIT itself. The role of magnetism in the semimetal phase is unresolved, and chiral spin textures are reported to persist into the metallic regime \cite{Machida2010}. Additionally, while \textbf{q}=0 AIAO order coincides with the opening of a charge gap \cite{Witczak2013}, the MIT shows evidence of both Mott and Slater (or mean-field) character \cite{Nakayama2016}. As an alternative to bandwidth control of the MIT with A-site isoelectronic substitution, filling control via carrier doping has the potential to access metallic states in the A$_2$Ir$_2$O$_7$ phase diagram with thermodynamically distinct magnetic and electronic properties.  

Previous studies of filling control in the pyrochlore iridates via hole-substitution have shown varied magnetic responses. In recent work on $($Eu$_{1-x}$Ca$_x)_2$Ir$_2$O$_7$, the MIT and AIAO transition temperatures remain coincident \cite{Kaneko2019}, and this transition is rapidly suppressed with increasing $x$ until a metallic ground state is realized between $x$=0.05 and $x$=0.10. However, other studies on A=(Y,Ca) and A=(Eu,Sr) describe different behavior entirely: while the MIT temperature decreases rapidly with $x$, the magnetic transition is decoupled in temperature, either fixed to the $x$=0 value with another fixed high-temperature transition \cite{Zhu2014} or more slowly decreasing with $x$ \cite{Banerjee2017}. The discrepancies between these na\"{i}vely similar materials, and the relation between their synthesis conditions and magnetic properties, warrants further investigation. Specifically, how the interplay between structure and magnetism evolves upon carrier-doping and how the presence of magnetism on the A-site affects the evolution of the filling-controlled MIT remain open questions.

Here we study the effects of Ca substitution on the Nd-site of (Nd$_{1-x}$Ca$_x$)$_2$Ir$_2$O$_7$ with 0${\leq}x{\leq}$0.08 as a means of hole-doping across the MIT in the pyrochlore iridate phase diagram. Electron transport and magnetization measurements combined with X-ray diffraction (average and local structure) and X-ray absorption techniques are utilized to resolve the interplay between lattice, spin, and charge degrees of freedom across the MIT.  Diffraction measurements of the average structure reveal minimal changes in bond lengths and bond angles as Ca is alloyed into the pyrochlore matrix, and the local structure shows that Ca enters the lattice homogeneously with no resolvable clustering. As Ca substitution levels increase, the MIT is pushed downward in temperature and coincides with the onset of magnetic order.  For doping levels greater than $x=0.05$, the ground state switches to a metal with a weak upturn in the low temperature resistivity coupled to Nd magnetism. X-ray absorption spectroscopy shows that the metallic state retains a strong spin-orbit coupled character with branching ratios little altered from the undoped material, and magnetic circular dichroism data collected at the Ir $L$ edges reveal an anomalous, weak net moment that survives across the MIT.  Our data establish a complex interplay between magnetism and the formation of the metallic state in hole-doped Nd$_2$Ir$_2$O$_7$.

\section{Methods}

Polycrystalline samples of $($Nd$_{1-x}$Ca$_x)_2$Ir$_2$O$_7$ were synthesized by a solid-state reaction. Powders (99.99$\%$, Alfa Aesar) of Nd$_2$O$_3$, CaCO$_3$, and IrO$_2$ were mixed in stoichiometric ratios, ground and heated at 1073 K in an alumina crucible in air for 18 h. Next, the mixtures were ground, pressed into pellets at 300 MPa within an isostatic press, placed in alumina crucibles, and heated at 1273 K in air for 8 days with one intermediate grinding. This step was repeated at 1323 K $-$ 1373 K. For several samples, the remaining Nd$_2$O$_3$ was reacted by adding 4 mol$\%$ additional IrO$_2$ to the powder before sintering the pellet at 1373 K in an alumina crucible sealed in a quartz tube under vacuum for 8 days with an intermediate grinding. For these samples, the final pellet was sintered at 1173 K for 2 days in air.

Samples were characterized by synchrotron powder X-ray diffraction (XRD) measurements at Beamline 11-BM of the Advanced Photon Source (APS) at Argonne National Laboratory, and the patterns were refined using the TOPAS software package. Refinements of the data exhibit the expected pyrochlore phase as well as small ($<$1.5$\%$) impurity fractions of each of the reactants and Ir metal. Complementary chemical characterization was accomplished with a Rigaku ZSX Primus IV  wavelength-dispersive X-ray fluorescence (WDXRF) spectrometer, wherein pressed pellet samples were quantitatively analyzed against standards composed of unreacted powders of known stoichiometry.

Synchrotron total scattering data for pair distribution function (PDF) analysis were collected at Beamline 6-ID-D at the APS using powder taken from the same batches as the 11-BM samples. Sieved powders with ${<}$44 $\mu$m particle size were sealed into Kapton tubes using copper wire and epoxy in a He-filled glove-bag to provide a thermal exchange gas. The samples were measured in transmission using an area detector. The 2D data were integrated to 1D diffraction data utilizing the Fit2D software.\cite{Hammersley1996} Corrections to obtain $I(Q)$ and subsequent Fourier Transform with $Q$\textsubscript{max}$=$24 $\AA$ to obtain $G(r)$ were performed using the program PDFgetX2.\cite{Qiu2004} Analysis of the total scattering data was performed using the PDFgui software suite\cite{pdfgui} over the range 1.75 $\AA$ $-$ 10.0 $\AA$.

Magnetotransport measurements were carried out in a Quantum Design DynaCool Physical Property Measurement (PPMS) system. Cut portions of sintered pellets were mounted with GE varnish in a four wire configuration using silver paint to create contacts. Current was driven perpendicular to the applied magnetic field, and voltage was measured with a dc resistance bridge. Magnetization data were collected using polypropylene capsules containing 20 mg of powder, and measured with a vibrating sample magnetometer (VSM) within a DynaCool PPMS or a MPMS3 Quantum Design SQUID magnetometer.

X-ray absorption spectroscopy (XAS) measurements were performed at Beamline A2 at the Cornell High Energy Synchrotron Source (CHESS), and X-ray magnetic circular dichroism (XMCD) measurements were performed at Beamline 4-ID-D at the APS. Sieved powders with ${\approx}$5 $\mu$m particle size were prepared on layers of tape to achieve a uniform sample thickness corresponding to nearly two absorption lengths. Both measurements were collected at the Ir $L_{2, \,3}$ absorption edges ($2p_{1/2, \;3/2}\rightarrow5d$) in transmission geometry. At A2, the energy of the incident X-ray beam was selected using a double-crystal $\langle$111$\rangle$ diamond monochromator that was detuned to reject higher harmonics, and the absorption was detected with ion chambers. At 4-ID-D, the incident energy was selected using a double-crystal $\langle$111$\rangle$ Si monochromator and circularly polarized X-rays were generated in helicity-switching mode at 13 Hz using a diamond phase retarder. The absorption was detected using a diode with lock-in amplification \cite{Haskel2007}. To screen out spurious signal, XMCD measurements were repeated under $\mu_0H$=$+$5 T and $-$5 T with fields oriented parallel and antiparallel to the incident wave vector.

\section{Experimental Results}

\subsection{Average and local lattice structure measurements}

\begin{figure}
\subfigure{
\includegraphics[trim=22mm 11mm 26mm 19mm, clip,width=0.45\textwidth]{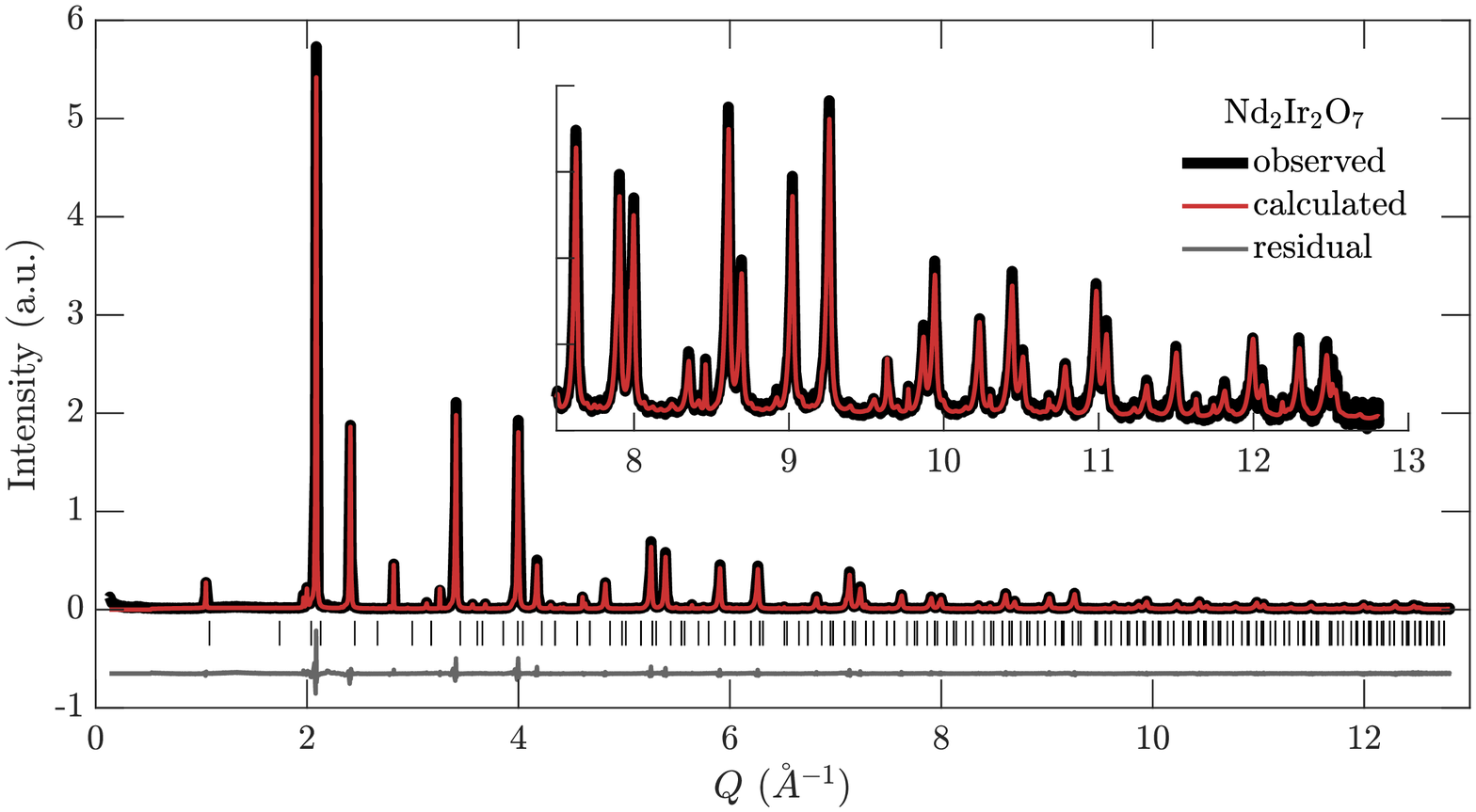}
\label{fig:XRDx0}
}
\subfigure{
\includegraphics[trim=22mm 11mm 26mm 19mm, clip,width=0.45\textwidth]{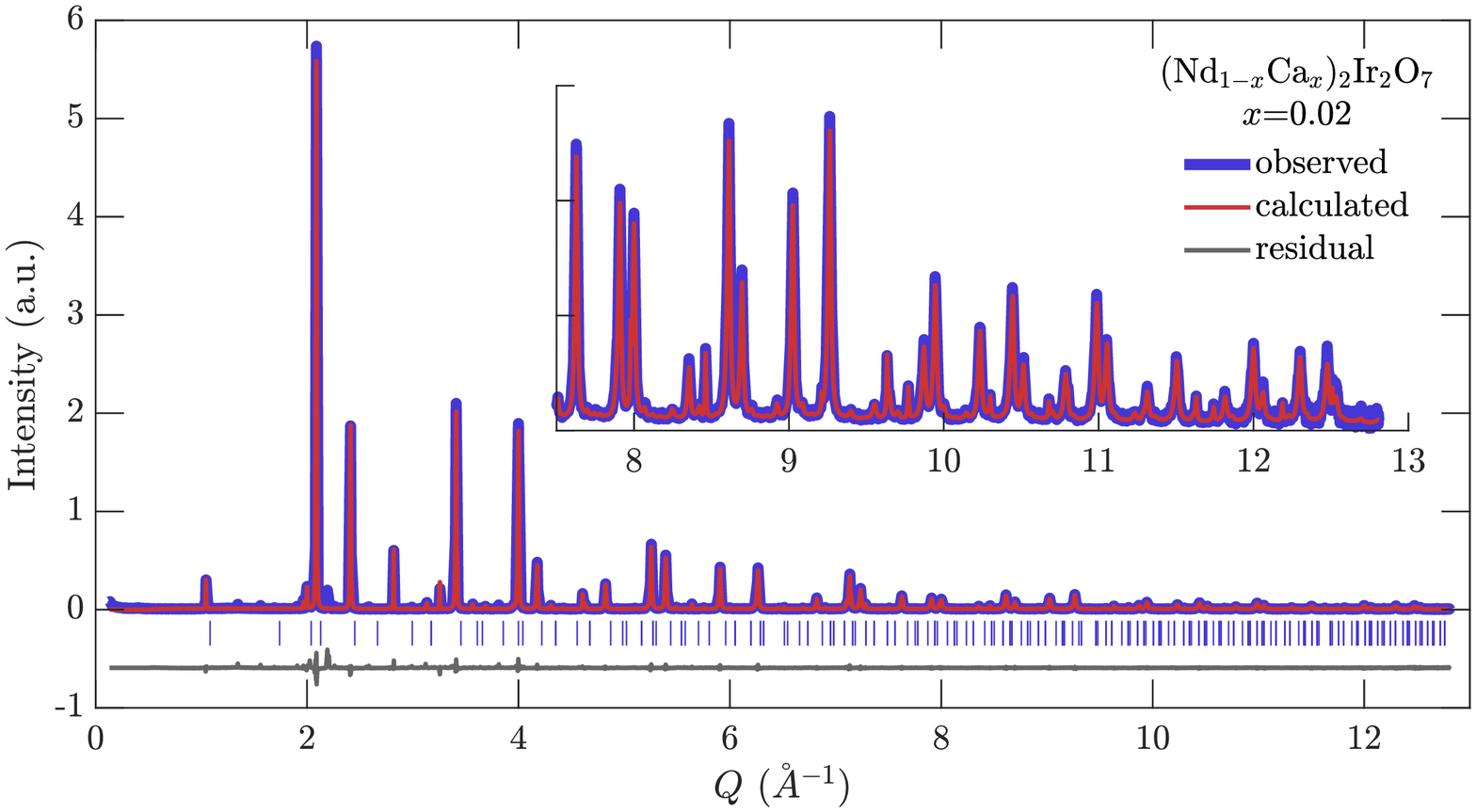}
\label{fig:XRDx2}
}
\subfigure{
\includegraphics[trim=22mm 11mm 26mm 19mm, clip,width=0.45\textwidth]{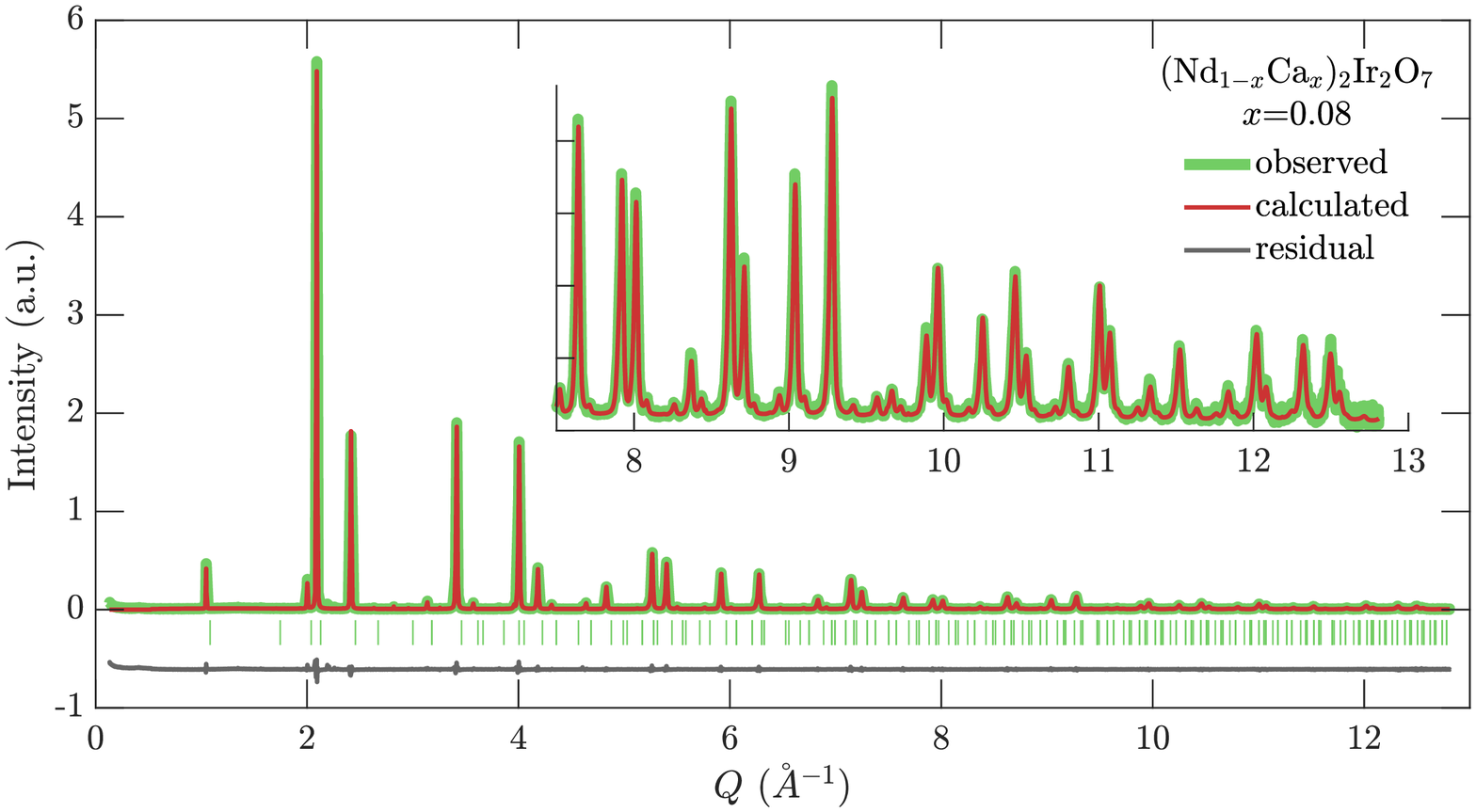}
\label{fig:XRDx8}
}
\caption{Synchrotron XRD patterns, all taken at 300 K. Calculated curves include the pyrochlore phase and the impurity phases Ir, IrO$_2$, and Nd$_2$O$_3$. Lines above the residual curves index only pyrochlore peaks.}
\label{fig:XRD}
\end{figure}

\newcolumntype{.}{D{.}{.}{8}}
\begin{table}
\setlength{\tabcolsep}{2pt}
\begin{tabular}{|l r|...|}
\hline
$x$ && 0 & 0.02 & 0.08 \\
$x$\tiny{\textsubscript{WDXRF}}	&& 0 & 0.0245(9)	& 0.0755(9)	\\
\hline
$a$(6 K) &[$\AA$]   & 10.3719(7) & 10.3667(9) & 10.3497(9) \\
$a$(25 K) &[$\AA$]  & 10.3712(9) & 10.3680(9) & 10.3497(9) \\
$a$(45 K) &[$\AA$]  & 10.3726(9) & $-$ & 10.3489(9) \\
$a$(300 K) &[$\AA$] & 10.3877(12) & 10.3783(12) & 10.3631(11) \\
\hline
\textbf{100 K} && & & \\
$u$(O$_{48f}$) && 0.3323(2) & 0.3325(3) & 0.3309(2) \\
$\angle$Ir-O-Ir &[$^{\circ}$] & 129.7(2) & 129.5(2) & 130.4(2) \\
$\angle$O-Ir-O  &[$^{\circ}$] & 82.28(7) & 82.16(8) & 82.76(7) \\
Nd       A Occ. && 1.000(16) & 0.975(17) & 0.925(11) \\
Ir $\;\;$A Occ. && 0.000(16) & 0.000(17) & 0.000(11) \\
Ir $\;\;$B Occ. && 0.988(23) & 1.000(25) & 1.000(17) \\
Nd       B Occ. && 0.012(23) & 0.000(25) & 0.000(17) \\
A $U_{iso}$ &[$\AA^2$]  & 0.00262(9) & 0.00305(9) & 0.00339(6) \\
B $U_{iso}$ &[$\AA^2$]  & 0.00344(7) & 0.00329(6) & 0.00228(4) \\
A$_2$B$_2$O$_7$ &[$\%$] & 96.48 & 95.59 & 98.39 \\
$R$\textsubscript{wp} &[$\%$] & 10.54 & 12.83 & 9.45 \\
$\chi^2$ && 2.32 & 3.00 & 2.24 \\
\hline
\textbf{300 K} && & & \\
$u$(O$_{48f}$) && 0.3313(2) & 0.3312(2) & 0.3298(2) \\
$\angle$Ir-O-Ir &[$^{\circ}$] & 130.6(2) & 130.1(2) & 131.1(2) \\
$\angle$O-Ir-O  &[$^{\circ}$] & 82.87(8) & 82.55(8) & 83.19(7)\\
Nd       A Occ. && 1.000(14) & 0.975(16) & 0.925(10) \\
Ir $\;\;$A Occ. && 0.000(14) & 0.000(16) & 0.000(10) \\
Ir $\;\;$B Occ. && 0.974(21) & 0.978(23) & 1.000(16) \\
Nd       B Occ. && 0.026(21) & 0.022(23) & 0.000(16) \\
A $U_{iso}$ &[$\AA^2$]  & 0.00660(9) & 0.00729(9) & 0.00749(6) \\
B $U_{iso}$ &[$\AA^2$]  & 0.00268(6) & 0.00272(6) & 0.00292(3) \\
A$_2$B$_2$O$_7$ &[$\%$] & 96.68 & 96.04 & 99.00 \\
$R$\textsubscript{wp} &[$\%$] & 10.01 & 12.12 & 8.69 \\
$\chi^2$ && 2.20 & 2.88 & 2.04 \\
\hline
\end{tabular}
\caption{Select crystallographic data from Rietveld refinement of synchrotron powder XRD data. First, nominal sample $x$ and sample $x$\tiny{\textsubscript{WDXRF}} \small from quantitative WDXRF analysis. Second, cubic lattice parameters $a$ at several temperatures. Next, refined values at 100 K and 300 K: $u$ for the O 48$f$ site; nearest-neighbor Ir-O-Ir bond angles; intra-tetrahedron O-Ir-O bond angles; occupancies of Nd and Ir on the A and B sites; isotropic atomic displacement parameters $U_{iso}$ for A- and B-sites; pyrochlore phase fractions less Ir, IrO$_2$, and Nd$_2$O$_3$; and Rietveld goodness-of-fit parameters $R$\textsubscript{wp} and $\chi^2$. Note oxygen occupancies and $U_{iso}$ were fixed to 1 and 0.001, respectively.}
\label{table:XRD}
\end{table}

\begin{figure}
\subfigure{
\includegraphics[trim=13mm 1mm 14mm 12mm, clip,width=0.45\textwidth]{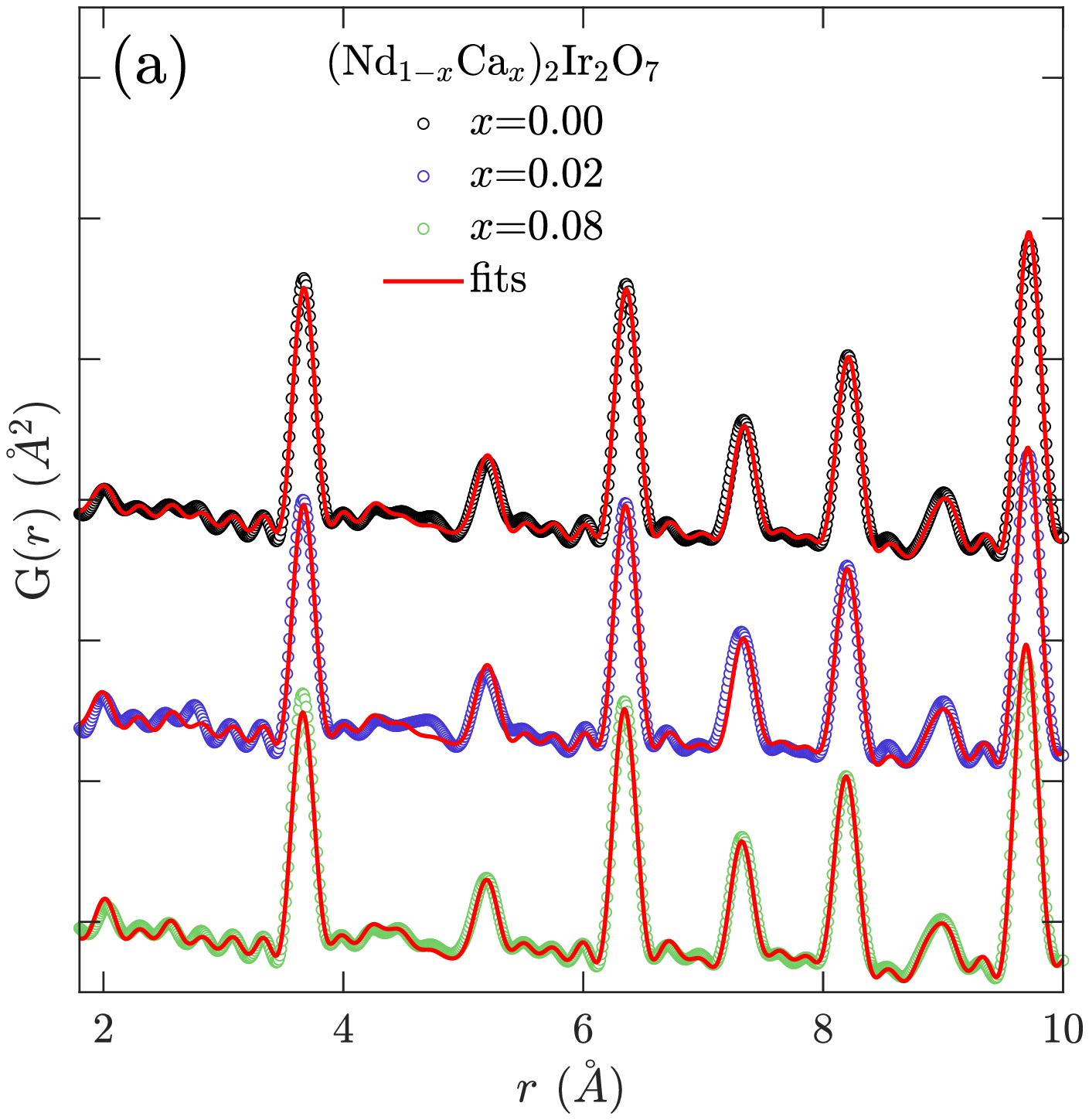}
}
\subfigure{
\includegraphics[trim=2mm 1mm 9mm 8mm, clip,width=0.45\textwidth]{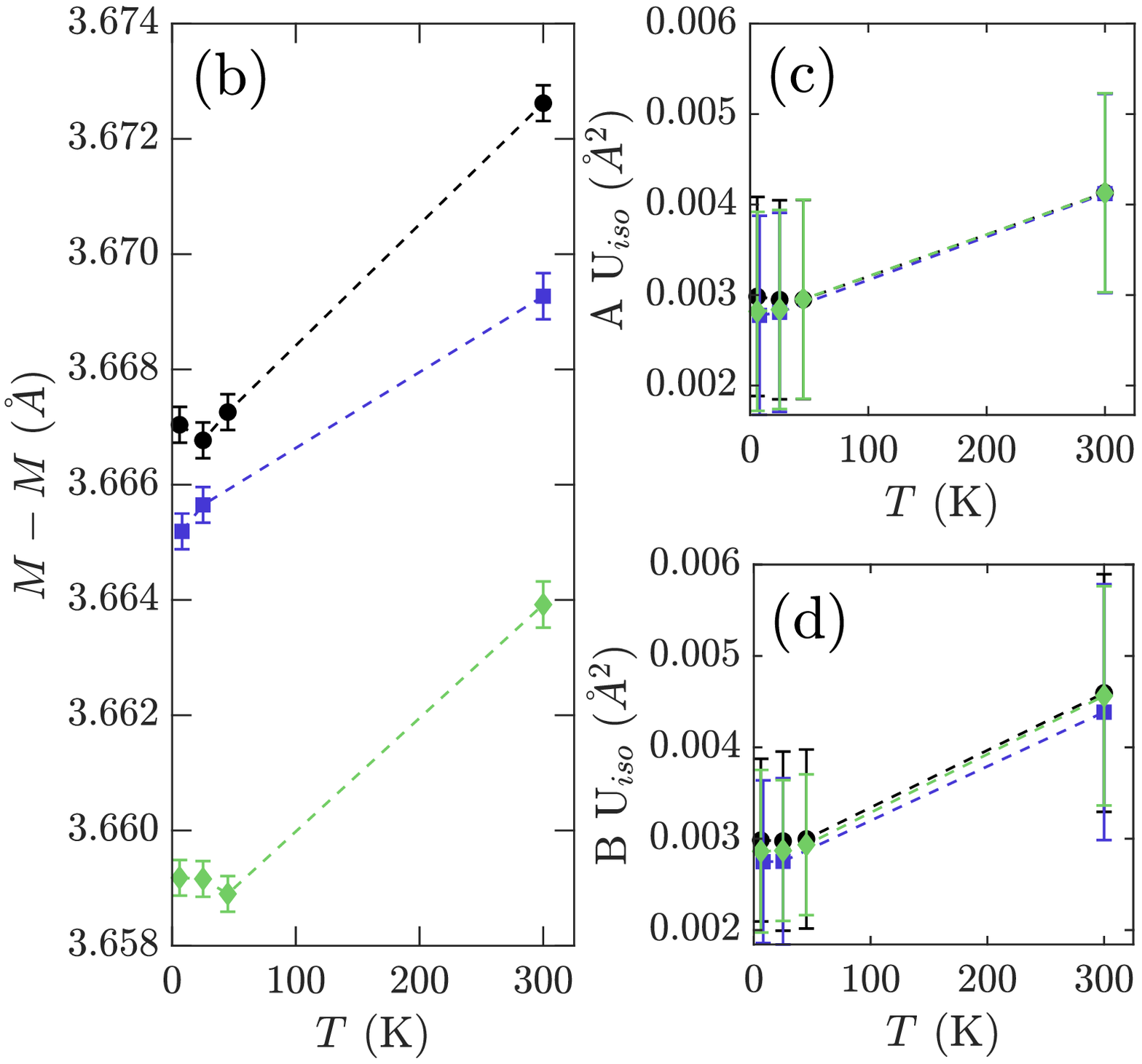}
}
\caption{X-ray PDF refinement of the A$_2$B$_2$O$_7$ local structure: (a) 300 K PDF data in circles, offset vertically, with fits in red solid lines. (b) Metal-metal distances between A and B sites. (c) A- and (d) B-site isotropic atomic displacement parameters $U_{iso}$. Dashed lines are guides to the eye.}
\label{fig:PDF}
\end{figure}

Synchrotron X-ray diffraction data were collected from samples across the doping range $0 \leq x \leq 0.08$ and powder patterns were indexed to the cubic space group $Fd\overline{3}m$. Bragg peaks were slightly asymmetric, indicative of a distribution of strains in the bulk which is commonly observed for pyrochlore samples \cite{Telang2018}. In the pyrochlore structure there are four sites A$_2$B$_2$O$_6$O$^{\prime}$, which are located at the $16d$, $16c$, $48f$, and $8b$ Wyckoff positions respectively. The cubic lattice constant $a$ decreases with Ca substitution (Table \ref{table:XRD}) in accordance with Vegard's law, as expected since 8-coordinate Ca$^{2+}$ ionic radii are nearly 1$\%$ smaller than Nd$^{3+}$. Scattering power from Ca and O sites was sufficiently weak that direct refinement of Ca and O occupancies was unreliable. Therefore, Ca concentrations were fixed at the WDXRF-measured values and fixed on the A-sites. This A-site preference for Ca is consistent with the small increase in atomic displacement parameters on the A-site with Ca content. Refinement attempts to place Ca on the B-sites resulted in inferior fits.

Occupancies refined for the A- and B-sites reflect a slight deviation from ideal stoichiometry. The XRD refinements indicate less than 3$\%$ `stuffing' through anti-site defects of excess Nd cations on the B-sites. 

The free coordinate $u$ for O $48f$ sites decreases marginally with Ca content. For the IrO$_6$ octahedra, this trend signifies a reduction in trigonal compression toward octahedral symmetry ($u{=}0.3125$). This reflects a small increase in the Ir-O-Ir bond angle, which changes at 300 K from 130.6(2)$^{\circ}$ in the $x$=0 sample to 131.1(2)$^{\circ}$ in the $x$=0.08 sample. This remains less than the $\approx$132$^{\circ}$ bond angle in the metallic A=Pr system \cite{Millican2007}, which is the value associated with the onset of metallicity in the bandwidth-driven global MIT of the A$_2$Ir$_2$O$_7$ series.  So while the increase in bond angle observed in the $x$=0.08 system implies a sterically-driven increase in bandwidth of the valence band due to enhanced Ir-O orbital overlap, the small bond angle change alone may not account for metallicity in the substitution-driven MIT for (Nd$_{1-x}$Ca$_x$)$_2$Ir$_2$O$_7$.

The evolution of the local structure was also analyzed via X-ray PDF experiments (Fig. \ref{fig:PDF}). No new atomic correlations were observed as Ca was introduced into the lattice or as the lattice was cooled through the MIT. Short-range order of the A- and B-sites is unchanged within experimental resolution for the measured samples for 6 K $\leq T \leq$ 300 K. All PDF measurements fit well to the site symmetry of the parent pyrochlore structure, with goodness of fit $R_w$ values between 9$\%$ and 15$\%$. This is consistent with a previous synchrotron XRD study that reported no symmetry change for Nd$_2$Ir$_2$O$_7$ upon cooling to 4 K \cite{Takatsu2014}. 

Notably, the A- and B-sites' isotropic displacement parameters, as determined from PDF refinements (Fig. \ref{fig:PDF}(c,d)), are also unchanged under varying Ca content within uncertainty. This precludes Ca clustering effects or nanoscale chemical phase separation. Furthermore, metal-metal distances track well with lattice constants, indicating that minimal site disorder is introduced with Ca substitution. 

\subsection{Magnetotransport measurements}

\begin{figure}
\subfigure{
\includegraphics[trim=5mm 3mm 14mm 8mm, clip,width=0.42\textwidth]{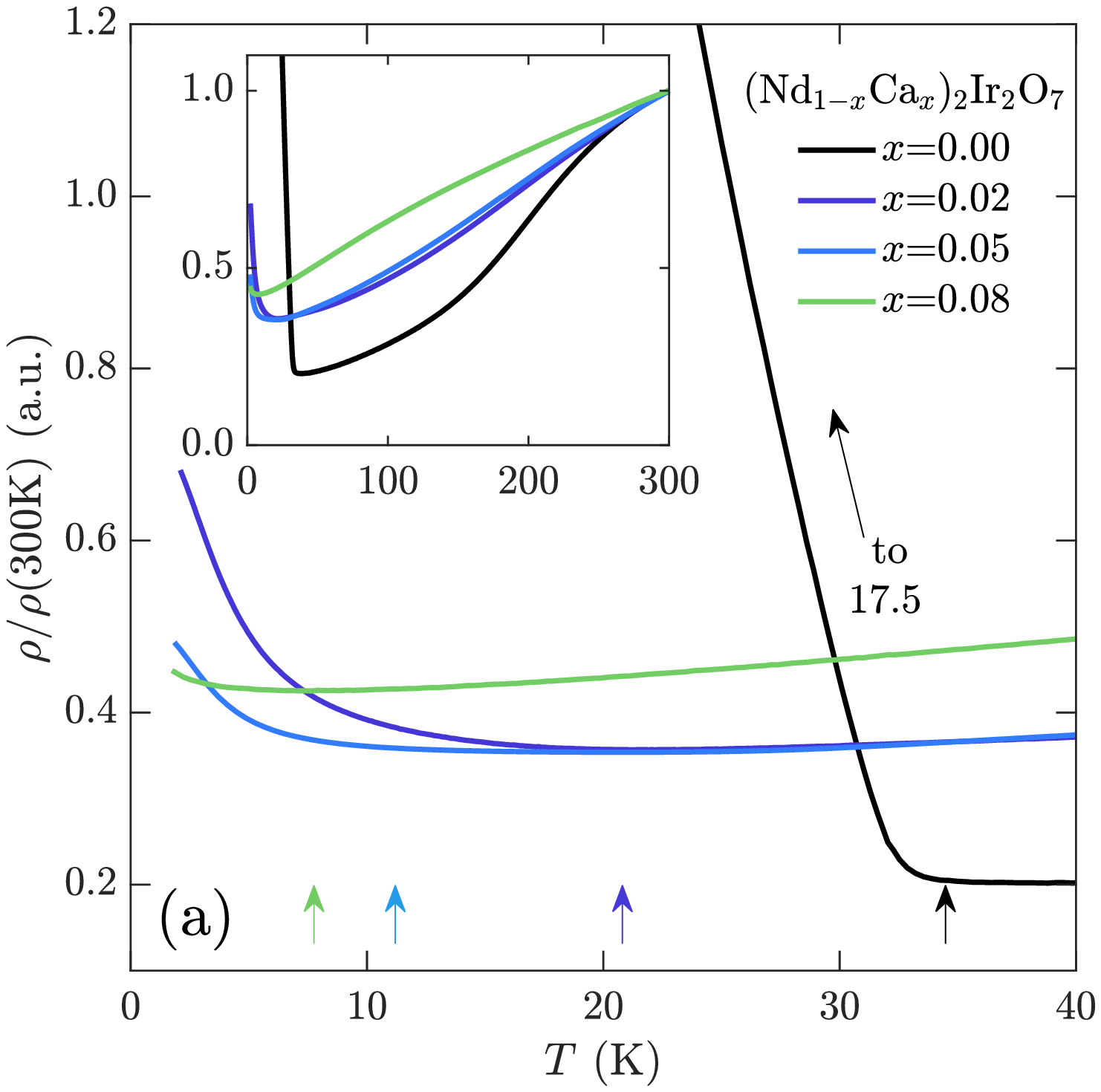}
\label{fig:res}
}
\subfigure{
\includegraphics[trim=5mm 2mm 14mm 12mm, clip,width=0.42\textwidth]{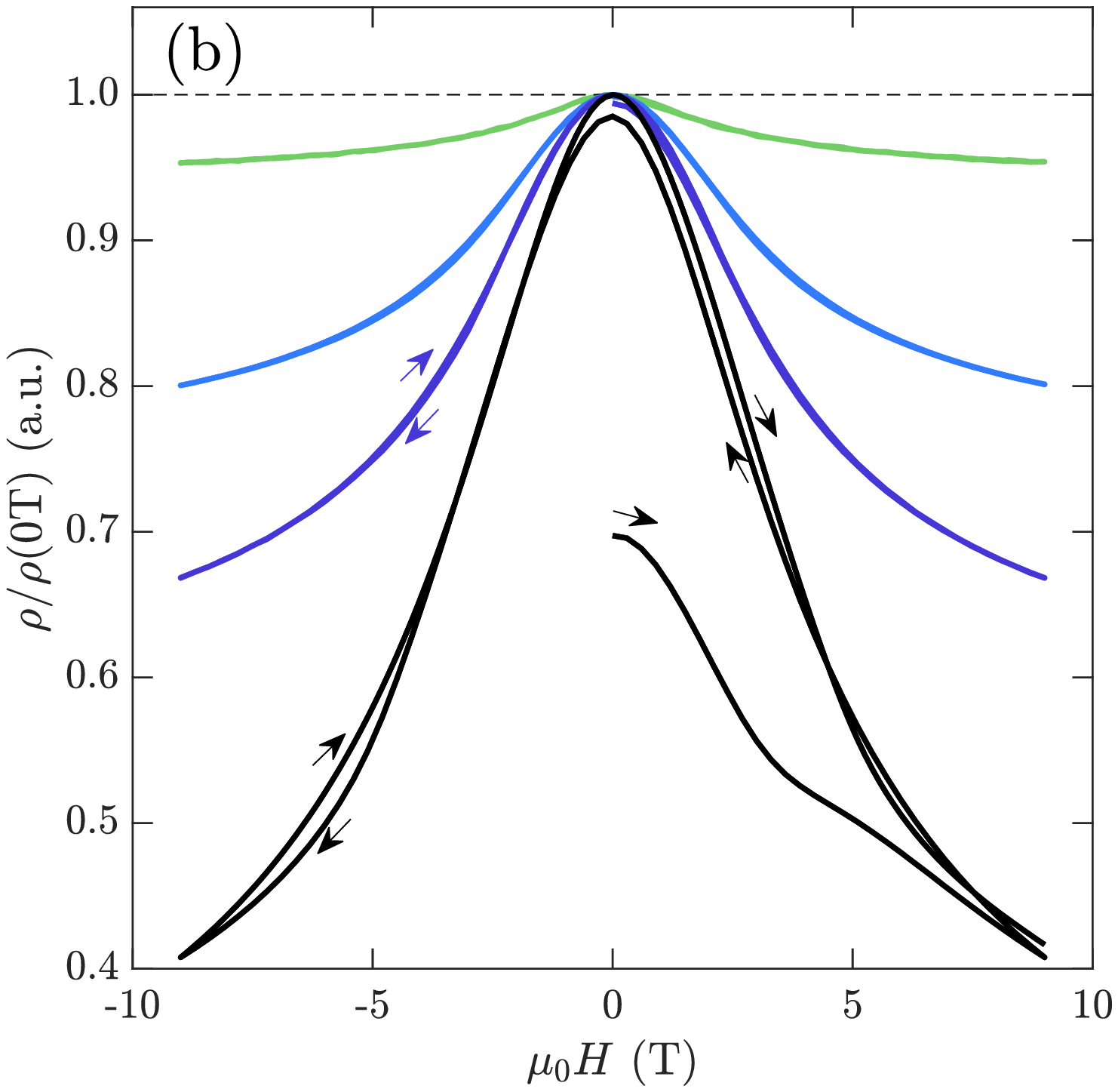}
\label{fig:MR}
}
\subfigure{
\includegraphics[trim=2mm 43mm 1mm 26mm, clip,width=0.43\textwidth]{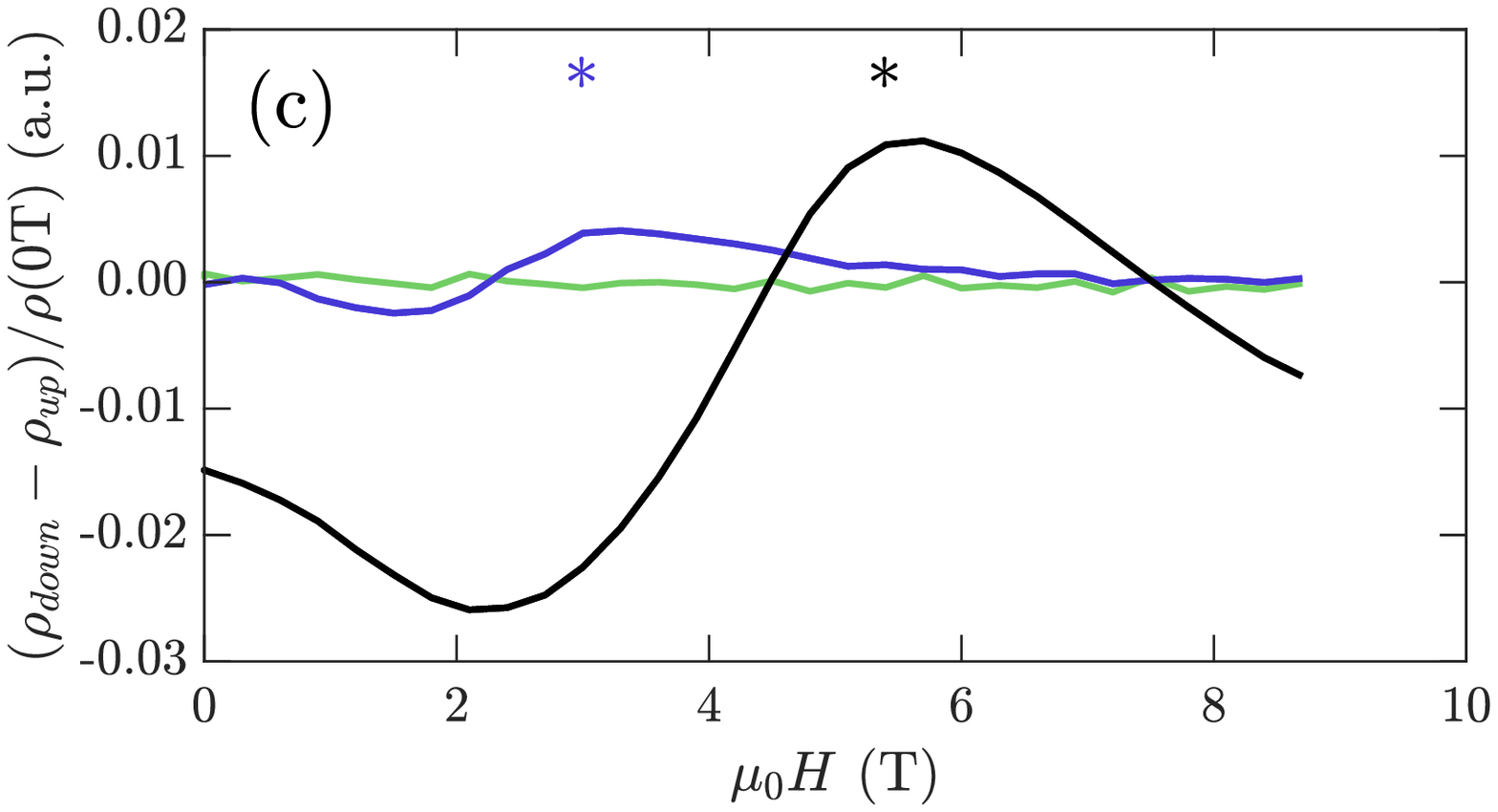}
\label{fig:MRdiff}
}
\caption{ Magnetotransport measurements. (a) Relative resistivity from 2 K to 300 K, all at 0 T on zero field cooling. Arrows indicate the onset of $\frac{\delta\rho}{\delta T}>0$. Resistances are normalized due to variation in pellet densities; typical values are $\rho($300 K$){\approx}10$ m$\Omega{\cdot}$cm. (b) Relative magnetoresistance at 2 K on zero-field cooling, swept from $\mu_0H{=}0{\rightarrow}{+}9{\rightarrow}{-}9{\rightarrow}{+}9$ T, with arrows to indicate field sweep direction. Note the hysteretic splitting, which is largest for the $x$=0 sample. (c) Hysteretic differences of the magnetoresistance in b, for sweeps after the initial  $0{\rightarrow}{+}9$ T sweeps (virgin curves).}
\end{figure}

\begin{figure}
\includegraphics[trim=0mm 0mm 0mm 1mm, clip,width=0.45\textwidth]{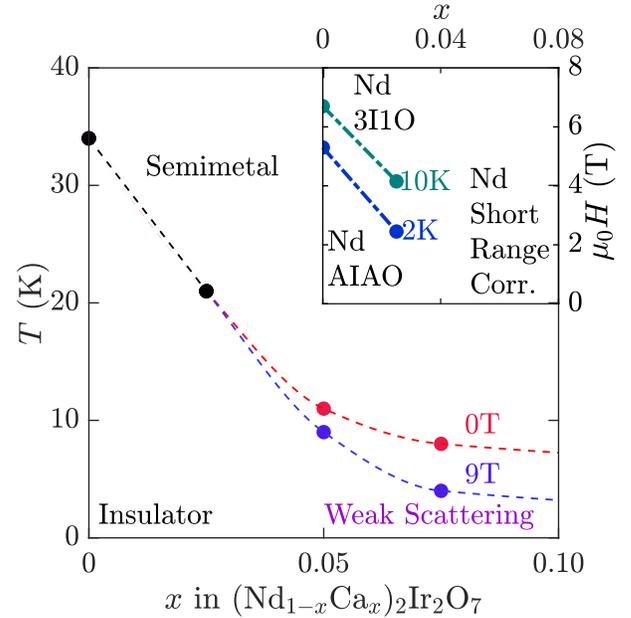}
\caption{Temperature-concentration phase diagram based on resistivity measurements. Inset: field-concentration phase diagram, indicating magnetism on the Nd sublattice. All lines are guides to the eye.} \label{fig:phasediagrams}
\end{figure}

Resistivity measurements shown in Fig. \ref{fig:res} reveal high temperature metallic behavior in all (Nd$_{1-x}$Ca$_x$)$_2$Ir$_2$O$_7$ samples with a low temperature transition into an insulating state (defined via the slope $\frac{\delta\rho}{\delta T}$). The temperature of the resulting MIT monotonically decreases as a function of Ca content; but the ``insulating" state for samples with $x{\geq}$0.05 becomes ill-defined. Notably, the sharp $T$\textsubscript{MIT}$=$34 K transition in the parent ($x$=0) sample weakens and broadens with the addition of carriers. For samples with $x{\geq}$0.05, the weak residual upturn in resistivity is further suppressed with the application of a transverse magnetic field, as depicted in the phase diagram in Fig. \ref{fig:phasediagrams}.  This weak, remnant upturn in $\rho(T)$ upon cooling likely arises from carriers coupling to the residual Nd moments in the samples, and the upturn temperatures in the high-Ca $x{\geq}$0.05 samples match muon spin resonance ($\mu$SR) features linked to ordering or freezing of the Nd magnetic sublattice \cite{Disseler2012, Guo2013}. 

In contrast, the MIT in the low-Ca samples $x{<}$0.05 is unaffected by applied fields up to 9 T, consistent with the stronger effective mean field of the Ir sublattice \cite{Savary2014}. This is confirmed by the (Eu$_{1-x}$Ca$_x$)$_2$Ir$_2$O$_7$ system, a nonmagnetic A sublattice analogue, wherein the MIT and Ir sublattice antiferromagnetism are fully suppressed below 2 K at a similar value of $0.05{<}x{<}0.10$ \cite{Kaneko2019}.  We speculate, then, that the full $T$\textsubscript{MIT} suppression to 0 K occurs for $x{>}$0.05 but is obscured by low-temperature scattering associated with Nd magnetic order.

Magnetoresistance (MR) data are shown in Fig. \ref{fig:MR} and are negative at low temperatures for all samples. The low temperature MR decreases in magnitude with increased Ca substitution, consistent with the suppression of Ir magnetism in the system. Field-induced `training' or hysteretic behavior is observed in the parent system and is consistent with previous studies of conducting domain walls \cite{Tian2016, Ma2015} in the parent material; yet this behavior is rapidly suppressed with Ca-doping. At low fields $\mu_0 H<2$ T, the magnetoresistance remains nearly quadratic in the moment extracted from magnetization $\rho(H){\propto}M^2$ as seen in Ref. \onlinecite{Disseler2012} . At larger fields $\mu_0 H>3$ T the hysteresis in MR ascribed to a field-induced magnetic phase transition remains resolvable up to $x=0.02$. Both of these features are discussed further in the next section.

\subsection{Magnetization measurements}

\begin{figure}
\subfigure{
\includegraphics[trim=3mm 2mm 4mm 12mm, clip,width=0.45\textwidth]{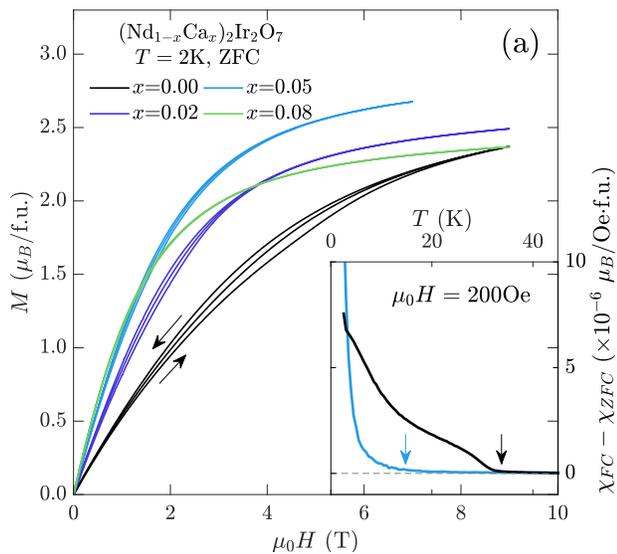}
\label{fig:MvH_MvT}
}
\subfigure{
\includegraphics[trim=5mm 1mm 13mm 6mm, clip,width=0.45\textwidth]{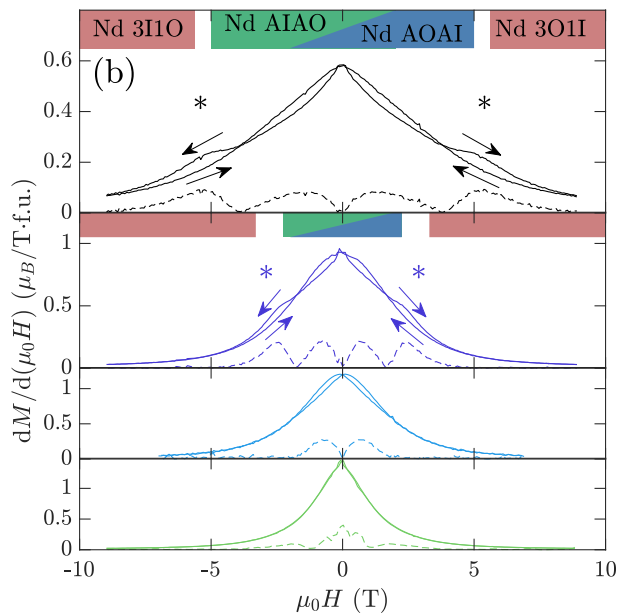}
\label{fig:MvHslope}
}
\caption{Magnetization measurements: (a) Field dependence of the isothermal dc magnetization, swept  as in Fig. \ref{fig:MR}, with just $H{>}0$ shown. Arrows indicate field sweep direction. Inset: Irreversibility of the dc susceptibility (field cooling minus zero-field cooling) taken at 200 Oe. $T$\textsubscript{MIT} values are indicated with arrows. (b) Solid lines are numerical derivatives of magnetization with respect to field, highlighting the hysteretic splitting. The initial  $0{\rightarrow}{+}9$ T sweeps (virgin curves) are not shown. Dashed lines are the absolute differences between sweep directions, magnified for clarity.}
\label{fig:MvH}
\end{figure}

Magnetization measurements shown in the inset to Fig. \ref{fig:MvH_MvT} reveal a weak irreversibility, hereby defined as the difference between field cooling (FC) and zero-field cooling (ZFC) sweeps, that persists across the doping range.  This is conventionally ascribed to either domain wall formation or spin canting within the all-in-all-out networks of Ir and Nd spins. Irreversibility in magnetization data appears at the same temperature where the low temperature resistivity changes slope. This connects the onset of magnetic correlations or freezing with the onset of the MIT for the low doping regime ($x{\leq}0.02$) as well as with the low temperature increase in resistivity for the high doping regime ($x{\geq}0.05$). We note here that in samples which were \emph{not} sintered under vacuum, an additional weak splitting between FC and ZFC data occurs near 120 K (Fig. \ref{fig:MvT_vac}), as was previously reported in the parent A=Nd system \cite{Disseler2013}. This is a synthesis-dependent effect with no resolvable influence on structural properties or $T$\textsubscript{MIT} or qualitative features in electron transport. We discuss this further in the Appendix.

Field-dependent magnetization data for the $($Nd$_{1-x}$Ca$_x)_2$Ir$_2$O$_7$ series are plotted in Fig. \ref{fig:MvH_MvT}. At $T=2$ K, magnetization data are expected to be dominated by the Nd sublattice \cite{Tian2016, Matsuhira2011}, and hysteretic differences appear between sweeps of increasing and decreasing fields. The splitting is illustrated by  plots of $\frac{\delta M}{\delta (\mu_0 H)}$ data shown in Fig. \ref{fig:MvHslope}. In the ordered state of the parent system, applying a magnetic field polarizes the Ising-like domains of both sublattices toward either AIAO or all-out-all-in (AOAI)\cite{Ueda2015a}. Upon a substantial increase in magnetic field, a second hysteretic feature appears, consistent with a spin-flop transition likely into the Nd 3-in-1-out (3I1O) state reported for the parent system \cite{Tian2016}. 

Upon Ca substitution, this higher field spin-flop feature decreases in onset field and vanishes for $x{=}0.05$. This is consistent with the destabilization of the Nd antiferromagnetic ground state at high hole-doping concentrations, and these features in magnetization mirror those in the magnetoresistance (Fig. \ref{fig:MRdiff}). The critical fields extracted from each measurement (marked by *) are equal within measurement uncertainty and suggest that the isothermal magnetoresistance data in Fig. \ref{fig:MR} is governed by domain scattering effects at low doping values and low temperatures.

\subsection{X-ray absorption spectroscopy and X-ray magnetic circular dichroism measurements}

\begin{figure}
\includegraphics[trim=1mm 13mm 1mm 2mm, clip,width=0.47\textwidth]{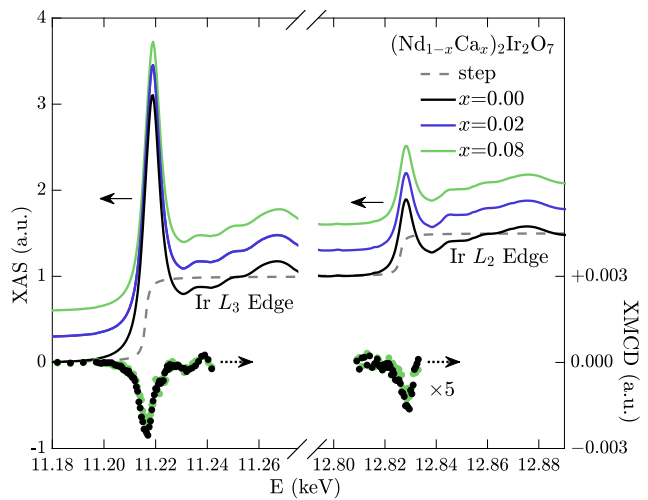}
\caption{X-ray spectroscopic measurements: Ir $L$-edge XAS is indicated with solid lines, and XMCD with dots. XAS lines are offset by 0.3 for clarity. XMCD was measured at 5 K and $\pm$5 T and normalized to the corresponding XAS edge jump.}
\label{fig:4IDD}
\end{figure}

\begin{table}
\centering
\setlength{\tabcolsep}{4pt}
\begin{tabular}{|c|cccccc|}
\hline
$x$ & $n_h$ & XAS $n_h$ & $BR$ & ${\langle}L{\cdot}S{\rangle}$  & $m_{tot}$ & $L_z/S_z$ \\
 & [$q_e$] & [$q_e$] & & [$\hbar^2$] & [$\mu_B/$Ir] & \\
\hline
0.00 & 5.00 & 5$\;\;\qquad$  & 5.7(2) & 2.8(1) & 0.0042(9) & 3.0(4)  \\
0.02 & 5.02 & 5.06(6) & 5.7(2) & 2.8(1) &$-$&$-$  \\
0.08 & 5.08 & 5.10(6) & 5.5(2) & 2.7(1) & 0.0036(9) & 2.8(4) \\
\hline
\end{tabular}
\caption{Calculations for $($Nd$_{1-x}$Ca$_x)_2$Ir$_2$O$_7$ from Ir $L$-edge XAS and XMCD, as described in the text: stoichiometric and XAS-calculated number of holes $n_h$; branching ratios $BR$; spin-orbit expectation values ${\langle}L{\cdot}S{\rangle}$; total moments $m_{tot}$ at 5 K and 5 T; and $L_z/S_z=2m_l/m_s$.}
\label{table:XAS}
\end{table}

XAS data were collected at the Ir $L_{2,3}$ edges (Fig. \ref{fig:4IDD}, Table \ref{table:XAS}) in order to probe the electronic structure of the $5d$ valence states. Qualitatively, the spectra for all samples are quite similar, with each showing nearly identical fine structure. This is consistent with the small structural changes and low carrier concentrations associated with the steric and valence modifications arising from $x{<}0.1$ levels of Ca substitution. It also supports the notion that the bandwidth is not significantly broadened via Ca-doping, either via steric effects or quenching of the orbital moment of the $t_{2g}$ manifold.

\footnotetext[1]{This ${\langle}T_z{\rangle}$ value differs from studies of nearly-isolated [IrCl$_6$]$^{2-}$ systems in the next reference (Pedersen et al. 2016), where ${\langle}T_z{\rangle}{\approx}0.5{\langle}S_z{\rangle}$, but note that the ligand and the bonding environment are different.}
XMCD measurements on all samples exhibit weak local Ir moments with resolvable signal just above the instrumental detection limit. Our application of the sum rules \cite{Thole1992} assumes a sizable magnetic dipole term ${\langle}T_z{\rangle}{\approx}0.2{\langle}S_z{\rangle}$ from Configuration Interaction calculations\cite{Note1, Pedersen2016} on several other IrO$_6$ systems BaIrO$_3$ \cite{Laguna2010} and Sr$_2$IrO$_4$ \cite{Haskel2012}. From the inclusion of ${\langle}T_z{\rangle}$, which slightly decreases the effective spin moment, we calculate the Ir total net moment $m_{tot}$=0.004 $\mu_B/$Ir at 5 K and 5 T, which is unchanged within uncertainty for the $x$=0 and $x$=0.08 samples. This magnitude is similar to the values 0.008 and 0.011 $\mu_B/$Ir reported from bulk dc magnetometry on the nonmagnetic A sublattice analogue systems A=Y \cite{Zhu2014} and A=Lu \cite{Yang2017}.

Both XAS and XMCD data indicate that the samples are near a $J$\textsubscript{eff}=$1/2$ state. The XAS $L_2$ and $L_3$ white line intensities are defined as $I_{L_{2,3}}=\int[\mu(E)-\Theta(E)] dE$, where $\mu(E)$ is the XAS signal, $\Theta(E)$ is a broadened step function centered on the inflection energy as expected for isolated ions, and the integration range is over the white line feature \cite{Clancy2012}. The measured branching ratios $BR=I_{L_3}/I_{L_2}$ are much higher than the statistical value for free ions of 2, which from the selection rules $\Delta j$=$0,\pm1$ indicates that unoccupied $5d$ states are primarily $5d_{5/2}$ rather than $5d_{3/2}$. Thus, $BR$ is a relative measurement of spin-orbit coupling for similarly prepared samples. The ratio $L_z/S_z{\approx}3$ from the XMCD sum rules \footnote{If we use the sum rules but instead, disregard the Configuration Interaction approximation and set ${\langle}T_z{\rangle}{=}0$, we find $L_z/S_z{\approx}$2 and $m_{tot}{\approx}0.005 \mu_B/$Ir.}, coupled with high $BR{\approx}6$, provides a strong indication of an Ir $J$\textsubscript{eff}=$1/2$ ground state \cite{Kim2008,Laguna2010}.

Now we comment more on hole doping and bandwidth changes from the spectroscopic studies. For the measured samples, the Ir $L_3$ XAS inflection and peak energies (of 11216.2 eV and 11218.8 eV respectively) only change within monochromator repeatability ${\approx}0.1$ eV, but the trend is increasing as expected for hole doping. More direct evidence of doping comes from the increase in the XAS-calculated number of holes, estimated from the change in  $I_{L_2}{+}I_{L_3}$ relative to the parent value in Table \ref{table:XAS}. Additionally, the XMCD peak energy is 0.4(2) eV higher for $x{=}0.08$ compared to $x{=}0$, corresponding to a lower $10Dq$ value for the photoexcited core-hole state.  While this is not a proper measurement of $10Dq$ for the ground state electronic configuration, which is typically ${\approx}$4 eV for the pyrochlore iridates,\cite{Hozoi2014} it suggests a small decrease in the $e_g{-}t_{2g}$ splitting with Ca substitution. 

\section{Discussion}

Substitution of Nd for Ca suppresses the MIT in (Nd$_{1-x}$Ca$_x$)$_2$Ir$_2$O$_7$ primarily via hole doping with relatively minor structurally induced changes to the bandwidth.  Once a doping level of $x$=0.08 is reached, signatures of Ir magnetic order vanish and transport behaves as a metal with weak disorder. We note that this value of $x$ for the paramagnetic metal phase is somewhat less than some theoretical predictions. Under hole doping alone, the collapse of AIAO order in the Y$_2$Ir$_2$O$_7$ is predicted by one study to occur for $n_h$=5.2 ($x$=0.20).\cite{Shinaoka2015} We reconcile this difference by considering the much weaker effective correlations $U/t$ in Nd$_2$Ir$_2$O$_7$.

A similar suppression of the MIT is observed upon doping holes directly onto the Ir-sites via B-site alloying.  In studies of polycrystalline Nd$_2($Ir$_{1-x}$Rh$_x)_2$O$_7$, Rh substitution was shown to strongly suppress $T$\textsubscript{MIT}=$T$\textsubscript{N} until a MIT was reached between $x$=0.05 and $x$=0.10 \cite{Ueda2012, Ueda2014}. Similar to the case of 8\% Ca-substitution in our study, for 10\% Rh-substitution both transport and magnetization reveal a weak upturn and irreversibility respectively near 5 K---likely arising from residual Nd moments freezing. The striking similarity between Rh and Ca substitution suggests comparable levels of hole-doping from each substituted cation. From a simplistic consideration of chemical potentials, Ir$^{4+}$ sites are expected to transfer electrons to the lower energy $J$\textsubscript{eff}=$1/2$ Rh$^{4+}$ states, forming the Ir$^{5+}$/Rh$^{3+}$ valence states observed in the IrO$_6$-based material Sr$_2$Ir$_{1-x}$Rh$_x$O$_4$ \cite{Clancy2014}. 

In comparison to carrier doping, external or ``chemical'' pressure (via introducing a larger A-site cation) modulating the bandwidth has a relatively gradual effect on the MIT. Both hydrostatic pressure and substitution of Nd$^{3+}$ for Pr$^{3+}$ yield a decrease in $T$\textsubscript{MIT}=$T$\textsubscript{N} with the suppression of the transition from chemical pressure proceeding gradually \cite{Ueda2015b}.  A future detailed structural study of Nd$_2$Ir$_2$O$_7$ under pressure, and in particular measurements of Ir-O bond lengths and angles in (Nd$_{1-x}$Pr$_{x}$)$_2$Ir$_2$O$_7$, would be insightful for resolving the precise role of Ca steric perturbations in assisting an MIT.

In the low temperature isothermal magnetization (Fig. \ref{fig:MvH}), the low field hysteresis for samples with $x{\leq}0.02$ likely arises from domain polarization of all-in-all-out and all-out-all-in magnetic domains, while at higher fields, the hysteresis observable only for $x{\leq}0.02$ is indicative of a field-induced spin-flop from Nd AIAO to 3I1O order.  This high field feature is consistent with the field identified in polycrystal \cite{Disseler2013} and single crystal studies on the parent Nd$_2$Ir$_2$O$_7$ compound \cite{Ueda2015a, Tian2016}. The disappearance of this feature for $x{\geq}0.05$ (i.e. for samples with a metallic ground state) suggests that long-range Ir correlations are quenched and the only remaining magnetization arises from short-range freezing of Nd/Ir moments coincident with the low temperature upturn in $\rho(T)$. In this picture, we assign the low field hysteresis for highly Ca-substituted samples to remnant short-range correlations primarily among Nd moments, as illustrated in the inset to Fig. \ref{fig:phasediagrams}.

Kondo coupling has been predicted to allow the alleged 3I1O Nd phase transition to be smooth \cite{Tian2016}. The temperature dependence of the resistivity in the insulating regime ($\frac{\delta\rho}{\delta T}{<}0$) reported here does not fit well to the empirical Hamann's expression of the Kondo effect $\rho{\propto}ln(T_K/T)$, unlike in another study of Pr$_2$Ir$_2$O$_7$ single crystals \cite{Nakatsuji2006}. This is likely attributable to the comparatively weaker Kondo coupling $J_K{\leq}10$meV predicted for the Nd system \cite{Tian2016}, which can be dominated by grain boundary-related scattering channels in polycrystalline samples. The quadratic low-field negative magnetoresistance is consistent with Kondo physics in doped samples; however future measurements on single crystals are required to fully explore this.

An open question remains regarding the origin of the weak, but finite, Ir ferromagnetism present in both the insulating parent $x$=0 and metallic $x$=0.08 samples. The small net moment extracted from XMCD at the Ir $L$ edges from both samples is identical within error and suggests that the weak local Ir moment is not trivially tied to the zero field long-range ordered state. Rather, the relative strength of Ir spins that couple to the applied field in both the insulating and metallic regimes is seemingly identical. We note here that the Ir XMCD signal may arise from either/both net ferromagnetism or a reversible response to the field. These cannot be distinguished without further measurements of the remnant magnetization obtained via hysteresis loops. Yet, if some of this signal is ferromagnetic, our measurements would be consistent with a picture of the weak ferromagnetism arising from antiferromagnetic domain walls in the parent insulating phase, wherein these domains persist locally in an electronically phase separated state well into the metallic regime. The fact that the polarized Ir moments at 5 T nearly match those observed in magnetization measurements of Y$_2$Ir$_2$O$_7$ and Lu$_2$Ir$_2$O$_7$ with nonmagnetic A-sites supports this notion and suggests a common origin to the weak ferromagnetism.

\section{Summary}

In summary, we report the suppression of both $T$\textsubscript{MIT} and the AIAO $T$\textsubscript{N} on the Ir sublattice via hole-doping within $($Nd$_{1-x}$Ca$_x)_2$Ir$_2$O$_7$. From a combined analysis of diffraction, PDF, and XAS data, we present evidence of Ca incorporation without clustering or phase separation on both local and average length scales. Calcium ions only weakly perturb the underlying structure with minimal changes inferred to the corresponding bandwidth, and hole carriers associated with replacing Nd$^{3+}$ with Ca$^{2+}$ cations instead drive the suppression of the low temperature MIT. For $x>0.02$, as the system enters a metallic ground state, both the charge transport and magnetism remain influenced by fluctuations and disorder on the Nd magnetic sublattice. Our results point toward the coincident suppression of long-range magnetic order and the charge gap in Nd$_2$Ir$_2$O$_7$ as the parent spin-orbit Mott state is suppressed via carrier doping. 


\appendix*  
\section{Effect of vacuum annealing on sample magnetization}

\begin{figure}
\includegraphics[trim=5mm 3mm 12mm 9mm, clip,width=0.45\textwidth]{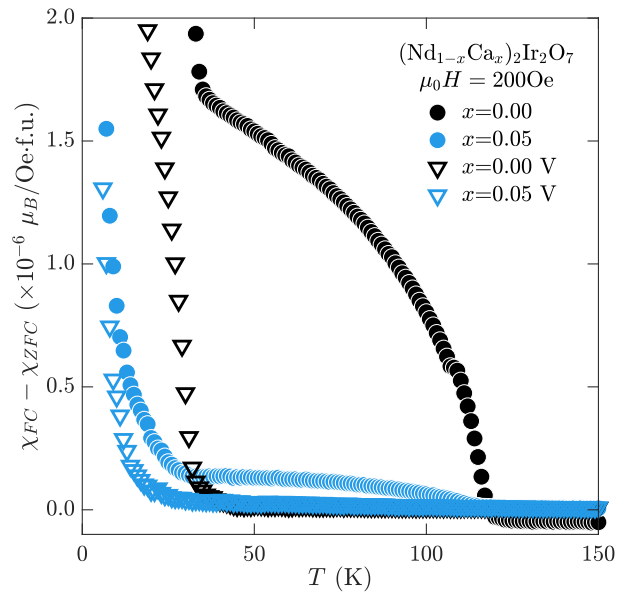}
\caption{Irreversibility of the dc susceptibility for samples before (circles, with features at 120 K) and after vacuum annealing (triangles, denoted `V').}
\label{fig:MvT_vac}
\end{figure}

We now discuss the observation of magnetic irreversibility near 120 K for samples of $($Nd$_{1-x}$Ca$_x)_2$Ir$_2$O$_7$ that were \emph{not} annealed in vacuum. We compare samples before and after vacuum annealing in Fig. \ref{fig:MvT_vac}, and attribute the additional feature to a secondary electronic phase within the bulk. We note that there is not a corresponding onset temperature in the Ir XMCD for samples that were not vacuum annealed.

A similar magnetic feature may be present in other pyrochlore iridates produced by solid state synthesis. In one study of $($Y$_{1-x}$Ca$_x)_2$Ir$_2$O$_7$ samples \cite{Zhu2014} there is one fixed transition temperature near $T$\textsubscript{MIT}($x$=0)=160 K and another that appears near 190 K for $x{>}$0. Since Y$^{3+}$ sites are nominally nonmagnetic, the weak magnetism likely arises due to the Ir sublattice. The origin of this variety of magnetic feature is not resolved by our study, but it may be related to clustering of nonmagnetic Ir$^{5+}$ or Ir$^{3+}$ sites. This hypothesis is supported by the measurable signal that is not attributable to Ir$^{4+}$ in the X-ray photoemission (XPS) spectrum for the Y$_2$Ir$_2$O$_7$ sample in the aforementioned study \cite{Zhu2014}. The likely source of Ir$^{3+}$ impurities is O$^{\prime}$ vacancies, a known issue in the defect-accommodating pyrochlore structure type \cite{Koo1998} which may scale with hole-doping \cite{Giampaoli2017}. 

\begin{acknowledgments}
This work was supported by ARO Award No. W911NF-16-1-0361 (S.D.W., Z.P., E.Z.). We thank Julian L. Schmehr and Ram Seshadri for helpful discussions. The research reported here made use of the MRL Shared Experimental Facilities, supported by the MRSEC Program of the NSF under Award No. DMR 1720256, a member of the NSF-funded Materials Research Facilities Network (www.mrfn.org). Additional support was provided by Bates College internal funding (G.L., S.B., S.H.). Research conducted at CHESS is supported by the NSF under Award No. DMR-1332208. Use of the Advanced Photon Source at Argonne National Laboratory was supported by the U.S. Department of Energy, Office of Science, under Contract No. DE-AC02-06CH11357.
\end{acknowledgments}

\bibliography{nd2ir2o7_resubmission}

\end{document}